\documentclass{aa}
\usepackage{graphicx}
\usepackage{txfonts}
\def \inte {\textit{INTEGRAL}}
\def \xmm {\textit{XMM--Newton}}
\def \sax {\textit{BeppoSAX}}
\def \src {4U\thinspace1850$-$087}
\def \glob {NGC\thinspace6712}

\def \nh {N${\rm _H}$}

\def \hcm {\hbox {\ifmmode $ cm$^{-2}\else cm$^{-2}$\fi}}

\begin{document}
   \title{The ultra-compact 
binary \src\ observed with \inte: hard X--ray emission 
from an X--ray burster\thanks{Based on
   observations with INTEGRAL, an ESA project with instruments and the science data
   centre funded by ESA member states (especially the PI countries: Denmark, France,
   Germany, Italy, Switzerland, Spain), Czech Republic and Poland, and with the
    participation of Russia and the USA.} }


   \author{L. Sidoli\inst{1}
                \and
      A. Paizis\inst{1}
          \and
          A. Bazzano\inst{2}
	  \and
	  S. Mereghetti\inst{1}
}

   \offprints{L. Sidoli, sidoli@iasf-milano.inaf.it}

   \institute{INAF-IASF, Istituto di Astrofisica Spaziale e Fisica Cosmica  Milano,
   via E.Bassini 15, I-20133 Milano, Italy
   \and
   INAF-IASF, Istituto di Astrofisica Spaziale e Fisica Cosmica Roma, 
   via del Fosso del Cavaliere 100, I-00133 Roma, Italy
          }
   
   \date{Received June 5, 2006; accepted September 2, 2006}


\abstract
  {The X--ray burster \src, located in the Galactic globular cluster \glob, 
is an ultracompact binary (orbital period $\sim$21 minutes), likely harbouring
a degenerate companion.}
   {The source has been observed at soft $\gamma$-rays several times with the \inte\ satellite, 
during the monitoring of the Galactic plane, with an unprecedented exposure time. 
We analysed all available \inte\ observations, with the main aim of studying the 
long-term behaviour of this Galactic bulge X--ray burster.}
  {The spectral results are based on the systematic analysis of all \inte\
observations covering the source position performed between March 2003 and November 2005.}
   {The source X--ray emission is hard and is observed, for the first time, up to 100~keV. 
A broad-band  spectrum obtained combining the  \inte\ spectrum 
together with a quasi-simultaneous  
\xmm\ observation performed in September 2003 is
well modeled with a disk-blackbody emission (with an inner disk temperature
of $\sim$0.8~keV) together with a power-law (with a photon index, $\Gamma$, of 2.1).
The 2--100~keV luminosity is 
1.5$\times$10$^{36}$~erg~s$^{-1}$ (assuming a distance of
6.8~kpc). 
}
  {\inte\ observations  reveal for the first time that 
this X--ray burster displays a very hard X--ray spectrum, with a cut-off
at energies higher than 100~keV, and that the source spends most
of the time in this low luminosity and hard state. 
Indeed, a previous \sax\ observation in April 1997 observed 
high energy emission from \src\ only up to 50~keV. 
}


\titlerunning{\inte\ observes \src\ up to 100 keV}
 
 \maketitle
%

\section{Introduction}

\src\ (Swank et al. 1976) is an X--ray burster
located in the galactic globular cluster
\glob. 
The likely optical counterpart displays an UV
modulation (Anderson et al.~\cite{a:93}) with a period of
20.6 minutes (interpreted as the orbital period), 
implying a degenerate companion of 0.04$M_\odot$ (Homer et al.~\cite{h:96}).

The first broad-band spectrum from this source 
was obtained with 
\sax\ in the energy range 0.3--50~keV during a pointing
performed in 1997 (Sidoli et al. \cite{s:01}).
The spectrum was well described with a disk-blackbody and Comptonized continuum with 
$N_{\rm H}$ = $3.9 \times 10^{21}$~\hcm, an inner disk temperature,
$kT_{\rm in}$, of $\sim$0.6 keV, an inner projected radius of $\sim$5~km
(for an assumed \glob\ distance of 6.8~kpc, Harris~\cite{ha:96}),
a temperature, $kT_{0}$, of the input ``seed'' photons, of 0.8$^{+0.1} _{-0.2}$~keV, an
electron temperature, $kT_{\rm e}$, of 70~keV, 
and an optical depth, $\tau$, of 1.7. 
The estimated 0.1--100~keV luminosity was 1.9$\times10^{36}$~erg~s$^{-1}$ (at 6.8~kpc).

Other X--ray observations are reported for \src\ limited
to the energy range below 10~keV ($ASCA$, Juett et al.~\cite{j:01};
\xmm, Sidoli et al.~\cite{s:05}; $Chandra$, Juett \& Chakrabarty~\cite{j:05}).

We report here the first detection of \src\ above 50~keV with the \inte\ satellite,
obtained with an unprecedented exposure time collected during the monitoring of
the Galactic plane.

\section{Observations and Results}

The ESA $INTEGRAL$ gamma-ray observatory, launched in October 2002,
carries three co-aligned coded mask telescopes: the imager IBIS
(Ubertini et al. 2003), which allows high angular resolution
imaging over a large field of view (29$^{\circ}\times29^{\circ}$)
in the energy range 15\,keV--10\,MeV, the spectrometer SPI
(Vedrenne et al. 2003; 20 keV--8\,MeV) and the X--ray monitor JEM-X
(Lund et al. 2003; 3--35\,keV).  IBIS is composed of a low-energy
CdTe detector (ISGRI; Lebrun et al. 2003), sensitive in the energy
range from 15\,keV to 1\,MeV, and a CsI detector (PICsIT; Labanti
et al. 2003), designed for optimal performance at 511 keV, and
sensitive in the 175\,keV--10\,MeV energy range.


We analyzed all public and Core Program IBIS observations pointed within
10$^{\circ}$ of the source. 
This resulted in 909 individual pointings (Science
Windows, SWs) performed between March 2003 and November 2005.
All the data have been processed using version 5.1 of the OSA \inte\ 
analysis software, and analysed with the corresponding rebinned response 
matrices.

A mosaic image of all 909 pointings revealed the sources that were active 
in this field during the period of interest.  The list of all the detected sources 
(detection significance $>$ 5$\sigma$) 
was then used in the spectral step, where spectra for all the 
detected sources in the field of view were simultaneously extracted with the 
standard spectral extraction method.

\vspace{1.5cm}
\begin{table}[ht!]
\caption{Summary of all \inte\ observations of \src\ analysed here.
The observations
have been grouped together in 4 data-sets for brevity.
The 3rd and 4th columns
list the Start and Stop Time
of the four groups of observations,
the fifth column reports the number of SWs in each data-set.
Data-set n.~4 includes Core Program data still not publicly available.
}
\label{tab:log}
\begin{center}
\begin{tabular}[c]{lllll}
\hline
\hline\noalign{\smallskip}
Data  & Temporal  & Start Time             &  End Time   &  Num of    \\
set          &  window & (MJD)                 & (MJD)          &  SWs    \\
\noalign{\smallskip\hrule\smallskip}
       1 &  Mar 2003--May 2003  &   52708.7  &        52772.0  &   258  \\
       2 &  Sep 2003--Nov 2003  &   52910.9  &        52963.4  &   210  \\
       3 &  Mar 2004--May 2004  &   53075.1  &        53126.5  &    84  \\
       4 &  Aug 2004--Nov 2005  &   53238.1  &        53684.0  &   357 \\
\noalign{\smallskip\hrule\smallskip}
\end{tabular}
\end{center}
\end{table}


 \begin{figure}
  \centering
   \includegraphics[angle=0,width=8.7cm]{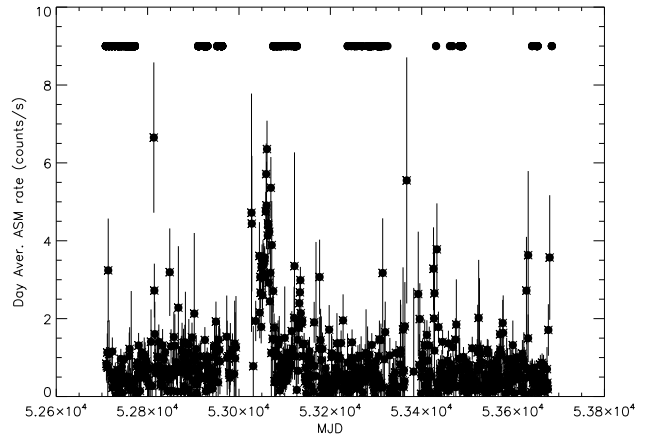}
      \caption{RXTE/ASM lightcurve (day average) of \src. The small solid circles
in the upper region show the times of the INTEGRAL observations analysed here.
              }
         \label{fig:asm}
   \end{figure}

\src\ is a relatively faint source, therefore 
a meaningful spectral analysis  needs to be performed
adding together several observations.
For this reason, we grouped the pointings in four data-sets (listed
in Table~\ref{tab:log}) from which we extracted 
four different IBIS/ISGRI spectra.

The 2--10 keV flux of \src\
shows a brightening event peaking on 25 February 2004, and lasting $\sim$10~days,
as visible from  the ASM/RXTE lightcurve reported in Fig.~\ref{fig:asm}.
Only part of its decay is covered by the INTEGRAL observations of data-set 3.

When we are averaging the spectra of \src\ using data that  
span over different mission times, we use a systematics of 5\% in the 
spectral fit process, to take into account for 
modifications in the 
instrument evolution (Lubinski et al., 2005).

The best-fit spectral results, using a single power-law 
are reported in Tab.~\ref{tab:4spec}
for the four data-sets.
The fits display a high $\chi$$^{2}$ (with a null hypothesis probability 
in the range 0.1--2.4\%) which can  be
explained by the spectral distortions
present when extracting IBIS/ISGRI spectra 
spanning very different mission times (Lubinski et al. 2005), especially for faint sources,
and by possible intrinsic source spectral changes on smaller timescales, 
which cannot be {\em a priori} excluded. Nevertheless, the global spectral shape is rather
constant on timescales of months--years.

\vspace{1.5cm}
\begin{table}[ht!]
\caption{Results of the spectral analysis of the four
temporal-selected IBIS/ISGRI spectra.
Data-set numbers are the same as in Table~\ref{tab:log}. 
The 2nd column reports the net exposure time of each spectrum. 
The 3rd column lists the IBIS/ISGRI average rates in the energy range
20--100~keV. Fluxes are in units of 10$^{-10}$~erg~cm$^{-2}$~s$^{-1}$(20--100 keV).
}
%
\label{tab:4spec}
\begin{center}
\begin{tabular}[c]{lrlllr}
\hline
\hline\noalign{\smallskip}
Data-set  & Exp. & IBIS/ISGRI rate &   Photon Index      &  Flux         &   $\chi^{2}$/dof   \\
          &  (ks)        & (s$^{-1}$)     &                     &  &             \\
\noalign{\smallskip\hrule\smallskip}
       1 &  372    & 1.25$\pm{0.05}$     & 2.5$\pm{0.2}$   &  0.8      &   21.6/8   \\
       2 &  383    & 2.36$\pm{0.05}$      & 2.1$\pm{0.1}$   &  1.8      &  18.8/7   \\
       3 &  114    & 1.42$\pm{0.09}$       & 2.2$\pm{0.3}$   &  1.1      &   11.6/7   \\
       4 &  490    & 1.59$\pm{0.05}$       & 2.2$\pm{0.1}$   &  1.2      &   16.1/7   \\
\noalign{\smallskip\hrule\smallskip}
\end{tabular}
\end{center}
\end{table}

We  tried also to extract JEM-X spectra, but the smaller field of view, 
the faintness of the source, together with 
some instrumental issues (e.g. JEM-X
switched-off in several occasions), severely reduced the
number of useful observations.
Thus data could be used only for the integrated 
spectrum covering the data-set 3 and part of the data-set 4.
The JEM-X spectra correspond to 76.7~ks net exposure time for JEM-X1
and 
55.5~ks for JEM-X2, with a source count rate of 
0.23$\pm{0.02}$ and 0.22$\pm{0.03}$, 
respectively in the two JEM-X units (5--20~keV).

\begin{figure}
  \centering
   \includegraphics[angle=-90,width=8.7cm]{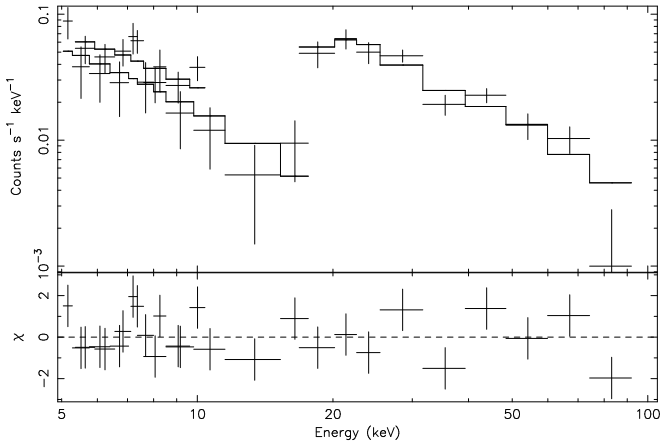}
      \caption{\src\ combined JEM-X and IBIS/ISGRI simultaneous spectrum,
fit with a single power-law (photon index of 2.2$\pm{0.2}$). 
In the lower panel, the residuals
in units of standard deviations are shown. 
              }
         \label{fig:spec}
   \end{figure}

The combined JEM-X plus IBIS/ISGRI simultaneous spectrum (taken from the data-set 3) is 
shown in Fig.~\ref{fig:spec}, fitted with an absorbed  
power-law, with absorbing column density fixed at 2.5$\times$10$^{21}$~cm$^{-2}$ (the
expected interstellar absorption towards the globular cluster \glob).
The fit is a good deconvolution of the spectrum ($\chi^2$/dof=27.9/23) and results in
a photon index of 2.2$\pm{0.2}$ and in a  5--100~keV flux corrected for the absorption
of (2.4$\pm{0.3}$)$\times$10$^{-10}$~erg~cm$^{-2}$~s$^{-1}$ 
(based on the IBIS/ISGRI response matrix). Other simple models,
like bremsstrahlung or a cut-off power-law, never result in a better fit.

\src\ has been
observed with \xmm\ on 27 September 2003 (see Sidoli et al. 2005), i.e. within the period covered
by \inte\ data-set 2, with the main aim
to study the low energy absorption intrinsic to the source. 
Indeed the \xmm\ spectrum below $\sim$2~keV is complex and we refer to Sidoli et al. 2005
for its detailed description (and for the \xmm\ data reduction). 
EPIC PN operated in Small Window mode, and with a net exposure time of 8.1~ks.
A combined spectral analysis EPIC/IBIS  interestingly 
allows us to extend the spectral study of this source in the soft X--rays,
below 5~keV, and to significantly refine the spectral parameters.
During the \xmm\ observation, 
only EPIC PN data did not suffer from pile-up problems,
thus we use this spectrum (1.7--12~keV, see Sidoli et al. 2005 for the
details) to perform a broad--band analysis of  the quasi--simultaneous 
EPIC--IBIS observations. 

\begin{figure*}
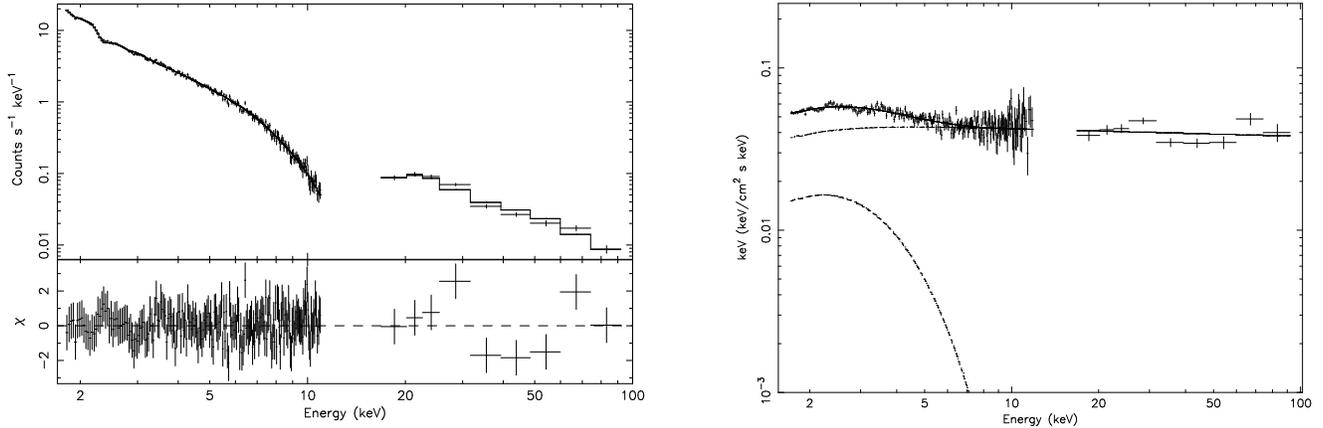

\hbox{\hspace{0.5cm}
\includegraphics[height=8.5cm,angle=-90]{5759fig3a.ps}
\hspace{1.0cm}
\includegraphics[height=7.7cm,angle=-90]{5759fig3b.ps}}
\caption[]{Broad-band \src\ EPIC/PN spectrum (obtained in September 2003) 
combined with the quasi-simultaneous IBIS/ISGRI (data-set 2). On the left, the 
count  spectra are displayed, together with residuals in units of standard deviations,
when fitted with a disk-blackbody plus a power-law.
On the right, the corresponding 1.7--100~keV  energy spectrum is shown.
}
\label{fig:broad}
\end{figure*}

Among the different models considered by Sidoli et al. (2005)
when fitting the \xmm\ spectrum, we tried their best-fit,
a disk-blackbody emission together with a power-law. 
This gives already a very nice result ($\chi^2$/dof=147.4/203), with the
spectral parameters reported in Table~\ref{tab:broad} and the spectrum shown
in Fig.~\ref{fig:broad}.
The normalization constant for the IBIS/ISGRI spectrum relative to PN
(constant factor fixed at 1), was 1.8$^{+0.2} _{-0.4}$. 
We note that this  normalization factor between the two instruments 
is in agreement with that  reported for the
Crab spectrum (Kirsch et al.~\cite{k:05}). 
The broad band flux (2--100~keV, corrected for the absorption)
is 2.8$\times$10$^{-10}$~erg~cm$^{-2}$~s$^{-1}$ 
(based on the EPIC PN response matrix).
Fixing the column density to the best-fit (5.7$\times$10$^{21}$~cm$^{-2}$)
found by Sidoli et al. (2005), does not
change the spectral results.

In order to obtain physical information, we then adopted 
the best-fit model used for the \sax\ \src\ spectrum (Sidoli et al. 2001), 
a disk-blackbody plus a Comptonized emission ({\sc compTT} model
in {\sc xspec}; Titarchuk \cite{ti:94}). 
Since the data extend only down to 1.7~keV,
we linked the temperature of the seed photons, kT$_{\rm 0}$, 
to the inner disk temperature, kT$_{\rm in}$. This is also justified
by the empirical finding that in ultra-compact binaries the two temperatures
are similar, within the uncertainties (Sidoli et al. 2001).
We obtain a good fit  ($\chi^2$/dof=156.3/203), 
with the parameters reported in Table~\ref{tab:broad}.
To obtain better constraints to the spectral paramenters, in this case 
the absorbing column density
has been fixed to the  \glob\ interstellar absorption value.
With this model, the normalization constant of the IBIS/ISGRI spectrum
with respect to EPIC PN lies in the range 1.5--1.9, again in agreement
with the standard value (Kirsch et al. 2005).
Adopting a cut-off power-law instead of a power-law
in the two-component model ($\chi^2$/dof=172.2/202), 
the cut-off energy resulted in E$_{c}$$>$110~keV (90\% confidence level), 
with a best-fit photon index of 1.9$\pm{0.1}$.

\vspace{1.5cm}
\begin{table*}[ht!]
\caption{Results of the spectral analysis of the broad-band
quasi-simultaneous spectrum (EPIC/PN--IBIS/ISGRI). 
Fluxes (2--100 keV) are corrected for
the absorption, are in units of 10$^{-10}$~erg~cm$^{-2}$~s$^{-1}$,
and are based on the EPIC pn response matrix. 
The assumed distance is 6.8~kpc (Harris et al. 1996). 
The {\sc compTT} model
assumes a spherical geometry. $i$ is the inclination angle of the disk. 
}
\label{tab:broad}
\begin{center}
\begin{tabular}[c]{lrlll}
\hline
\hline\noalign{\smallskip}
Model                     & \nh\              &   Parameters     &  Flux         &   $\chi^{2}$/dof   \\
                          &  (10$^{22}$~cm$^{-2}$)               &                  &               &                     \\
\noalign{\smallskip\hrule\smallskip}
{\sc diskbb}+{\sc pow}    & 0.4$\pm{0.2}$& $\Gamma$=2.07$^{+0.07} _{-0.15}$ & 2.8$\pm{0.1}$     &   147.4/203   \\
                          &                   & kT$_{\rm in}$=0.8$\pm{0.1}$ keV  &         &               \\ 
     &                    & r$_{\rm in}$$\times$$(cos(i))^{0.5}$=1.7$^{+1.1} _{-0.4}$  km &         &             \\ 
\hline
{\sc diskbb}+{\sc compTT} & 0.25 fixed             & kT$_{\rm e}$$>$40 keV  &  2.9$\pm{0.1}$     &  156.3/203   \\
                          &                    & kT$_{\rm 0}$=0.67$\pm{0.05}$ keV  &         &               \\ 
                          &                   & $\tau$=2.2$\pm{0.5}$             &         &               \\ 
                          &                   & kT$_{\rm in}$=kT$_{\rm 0}$     &         &               \\ 
     &                    & r$_{\rm in}$$\times$$(cos(i))^{0.5}$=4$\pm{1}$ km &         &             \\ 
\noalign{\smallskip\hrule\smallskip}
\end{tabular}
\end{center}
\end{table*}

\section{Discussion and Conclusions}

Previously, the source broad-band spectrum
was observed only with \sax\ in April 1997. At that time,
the PDS non-imaging instrument, covering the range from 20 to 200 keV, detected
the source only up to 50~keV (Sidoli et al. 2001).

We report here for the first time the discovery of hard (50--100~keV) 
X--ray emission from the X--ray burster \src\ and a long-term study of its X--ray 
spectral behaviour.
It is now possible to directly measure
the high energy luminosities of this source: they 
are $\sim$1.3$\times$10$^{36}$~erg~s$^{-1}$ and 7$\times$10$^{35}$~erg~s$^{-1}$
respectively in the 1--20~keV and 20--200~keV energy ranges (at 6.8~kpc). 
This implies that \src\ falls into
the so-called ``burster box'' (Barret et al., 2000).

In order to allow a proper comparison with the \inte\ spectrum
we re-analysed the \sax\ observation (April 1997; refer to Sidoli et al. 2001 
for the details of the data reduction) adopting the same model used here, 
a disk-blackbody emission together with a power-law.
This resulted in the following parameters ($\chi^2$/dof=170.4/158):
an absorbing column density of (0.46$\pm{0.03}$)$\times$10$^{22}$~cm$^{-2}$,
an inner disk blackbody temperature, kT$_{\rm in}$, of 0.66$\pm{0.03}$~keV,
and a powerlaw photon index of 1.96$\pm{0.06}$. 
Thus, there is a good agreement with the \inte\ spectroscopy.

\begin{figure}[ht!]
  \centering
   \includegraphics[angle=-90,width=8.0cm]{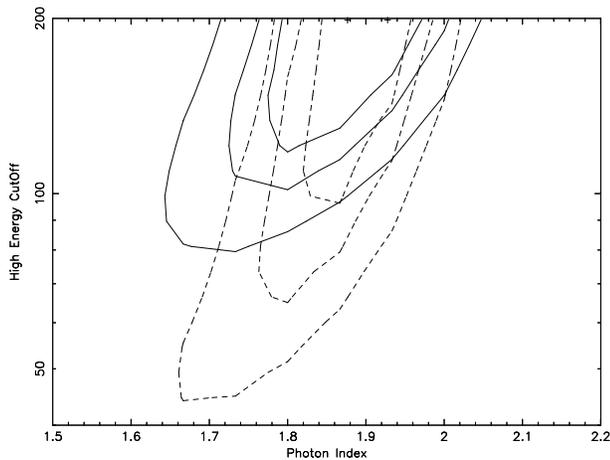}
      \caption{Comparison of the confidence contour levels for the
high energy cutoff (in units of keV): solid contours have been derived analysing EPIC--IBIS joint spectrum,
while the dashed contours mark the \sax\ results.
              }
         \label{fig:cutoff}
   \end{figure}

The new observations with \inte\ allow us to put much 
more stringent 
limits on the presence of a high energy cut-off.
In Fig.~\ref{fig:cutoff} we compare the confidence contour levels
for the high energy cut-off and the power-law photon index, obtained
with the EPIC--IBIS joint spectrum (solid contours) and
with \sax\ (dashed contours). 
During the \sax\
observation a lower limit to the high energy cut-off could be placed at E$_{c}$$>$60~keV (90\% level),
while now with \inte\ we can interestingly shift it towards much higher energies, with  E$_{c}$$>$100~keV.
This allows us to include \src\ among the few tens of low-luminosity 
low-mass X--ray binaries  (LMXRBs) for which the spectrum has been observed up to 100~keV (Barret et al. 2000,
Di Salvo \& Stella 2002, Bazzano et al. 2006). More interestingly, \src\ is now among the
hardest type I X--ray bursters in our Galaxy.

The \inte\ spectrum displays properties which fits well into the classification
of low luminosity ($\sim$0.01 L$_{\rm Edd}$) 
weakly magnetized neutron stars with ``hard spectra'' (Barret et al. 2001):
a broad-band spectrum extending up to 100~keV, 
well described by a soft disk-blackbody emission
together with a hard Comptonized component, with a optical depth of the Comptonizing 
corona of $\sim$2, seed photons temperature
below 1~keV, a high electron temperature, above 40~keV.
A few LMXRBs appear to spend most of the time in this ``hard state'' (Di Salvo \& Stella 2002),
like 4U~0614+091 (Piraino et al. 1999), which also displays spectral parameters similar
to \src, with a photon index in the range 2.3--2.4.
Also for \src\ the four different \inte\ spectra suggest that the source spend
most of its lifetime in this spectral state (on  timescales of months or years, if compared with \sax).

\begin{figure}[ht!]
  \centering
    \includegraphics[angle=-90,width=10.0cm]{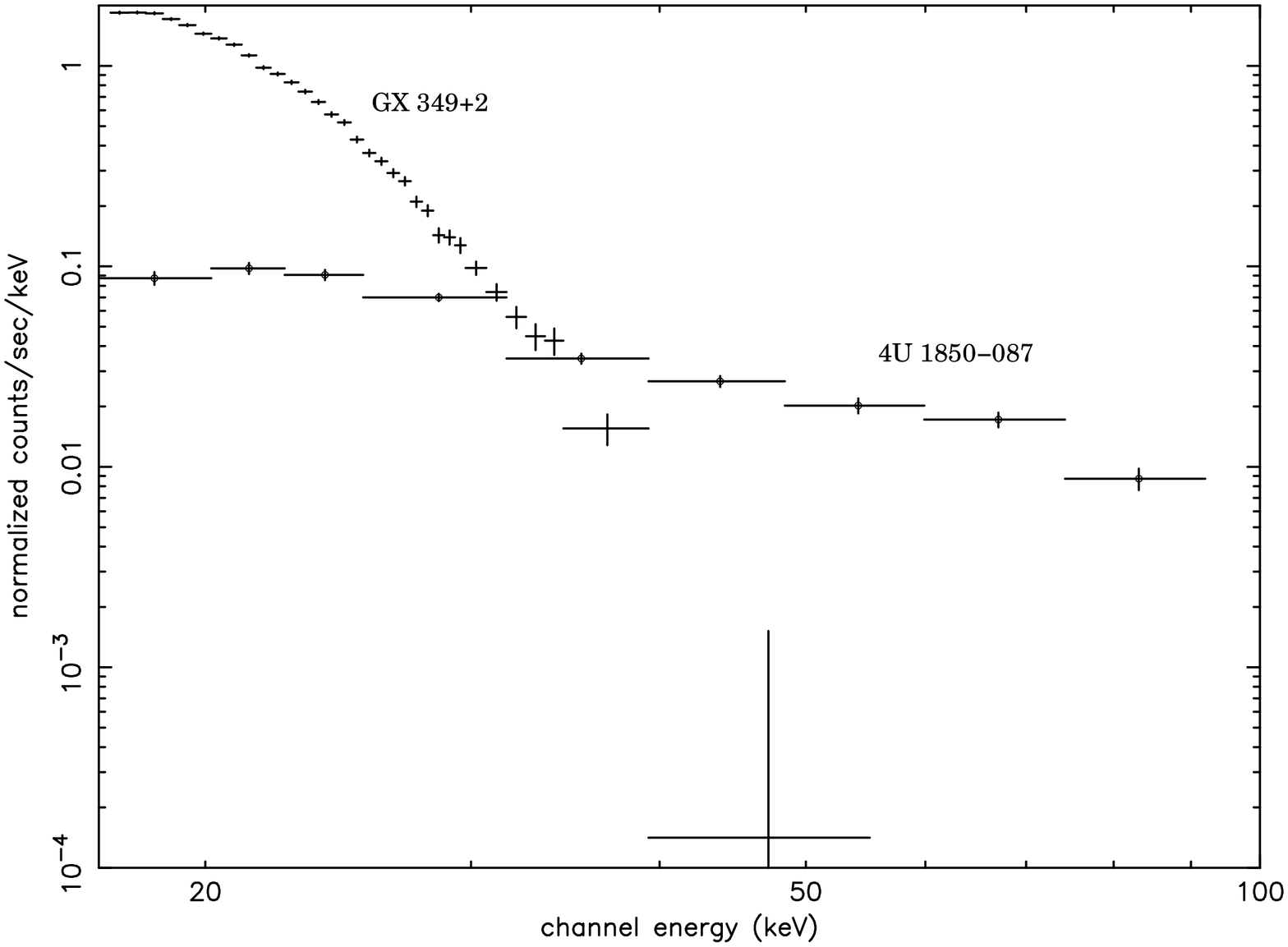}
     \caption{Comparison of the \inte\ spectrum of the source \src\ 
(data-set 2, the same dispayed
in Fig.~\ref{fig:broad}), with a 
spectrum of a typical Z-source, GX~349+2 (data from Paizis et al. 2006).
              }
         \label{fig:comp}
   \end{figure}

The very hard emission from \src\ is clearly evident also when compared (see Fig.~\ref{fig:comp})
to a typical spectrum of a Z-source, GX~349+2, 
reported in an \inte\ study of the spectral behaviour of the persistent LMXRBs 
hosting a neutron star (Paizis et al. 2006).

\begin{acknowledgements}
We acknowledge the Italian Space Agency financial and programmatic support
via contract I/R/046/04.
\end{acknowledgements}


\end{document}